\documentclass{appolb}

\usepackage{bm}
\usepackage{amsmath}
\usepackage{graphicx}
\usepackage{subfigure}
\usepackage[colorlinks=true,linktocpage=true,linkcolor=blue,citecolor=blue]{hyperref}
\usepackage[usenames,dvipsnames]{color}

\begin{document}

\title{Basic phenomenology for relativistic heavy-ion collisions
\thanks{email address: wojciech.florkowski@ifj.edu.pl}
}
\author{
Wojciech Florkowski
\address{The H. Niewodnicza\'nski Institute of Nuclear Physics, Polish Academy of Sciences, PL-31342 Krak\'ow, Poland, and \\
Institute of Physics, Jan Kochanowski University, PL-25406~Kielce, Poland}
}

\maketitle

\begin{abstract}
Basic concepts used to interpret the soft-hadronic data collected in relativistic heavy-ion collisions are reviewed at the elementary level. 
\end{abstract}

\section{Introduction}  

Physics of relativistic heavy-ion collisions is a very broad, interdisciplinary field of physics. It is impossible to cover its all important aspects in a short text.  Therefore, one has to decide which topics are selected for presentation. The idea behind these lectures is to give a possibly general overview which concentrates on soft-hadronic observables and may serve as the background for other advanced lectures presented during the School.

The omitted discussions are partly compensated by references given to many original papers. They will guide the reader in further studies. We also note that at present there are  several textbooks available which discuss heavy-ion collisions and physics of the quark-gluon plasma \cite{Csernai:1994xw,Wong:1995jf,Letessier:2002gp,Yagi:2005yb,Florkowski:2010zz}.  We refer to them and to the collected review articles \cite{Friman:2011zz} for additional information. 

In the remaining part of Introduction we give a historical outline of the development of the heavy-ion physics, present the main theoretical tools, and discuss the concepts of the quark-gluon plasma. In Sec.~\ref{sect:physters} we discuss the main physics terminology used in our field. Section \ref{sect:hag} is devoted to the concept of the limiting Hagedorn temperature. This concept has turned out to be very inspiring in the physics of strong interactions and led us directly to the idea of the phase transition from ordinary hadronic matter to quark matter (later called the quark-gluon plasma \cite{Shuryak:1978ij,Shuryak:1980tp}). In Sec.~\ref{sect:harmflow} we discuss shortly the coefficients of the Fourier expansion of the momentum distribution in the azimuthal angle. The first three coefficients are known as the directed, elliptic and triangular flows. The large values of the elliptic flow have been reproduced within the hydrodynamic calculations and suggest the small viscosity to entropy density ratio of the quark-gluon plasma. In Sec.~\ref{sect:glauber} we present the basics of the Glauber model, which is commonly used to determine initial distributions of the entropy or energy density in the transverse plane. The Glauber model serves also as a tool in comparisons between heavy-ion and more elementary proton--nucleus and proton--proton collisions.  The hydrodynamic approaches are discussed in Sec.~\ref{sect:hydro}, where we first recall the famous Fermi, Landau and Bjorken models and then switch to the characteristics of the perfect-fluid and viscous-fluid dynamics. The final stage of the space-time evolution of matter is shortly called a freeze-out. Several approaches to deal with this rather complicated process are presented in Sec.~\ref{sect:fout}. The space-time dimensions of the produced system at freeze-out can be inferred by the study of the identical particle correlations, which we shortly discuss in Sec.~\ref{sect:hbt}. The lectures are closed with short conclusions. Throughout  the text we use natural units where $c=\hbar=k_B=1$.

\subsection{Historical background}

The name ``heavy-ions'' is used for heavy atomic nuclei, and the term ``relativistic energy'' denotes the energy regime where the kinetic energy is much larger than the rest energy.  The first experiments with the relativistic heavy ions (energies larger than 10 GeV per nucleon in the projectile beam) took place at the  Brookhaven National Laboratory (BNL) and at the European Organisation for Nuclear Research (CERN) in 1986. The Alternating Gradient Synchrotron (AGS) at BNL accelerated ${}^{28}$Si at 14 GeV per nucleon. At CERN, the Super Proton Synchrotron (SPS) accelerated ${}^{16}$O at 60 and 200 GeV per nucleon in 1986, and ${}^{32}$S at 200 GeV per nucleon in 1987.  In 1990 the next project on heavy-ion physics was organised at CERN with ${}^{32}$S beams. In 1992 the experiments with $^{197}$Au beams at 11 GeV per nucleon were initiated at BNL. In 1995 the new experiments took place at CERN with ${}^{208}$Pb beams at 158 GeV per nucleon. 

In 2000 the first data from the Relativistic Heavy Ion Collider (RHIC) at BNL were taken. During the first run, the maximum energy of 130 GeV per nucleon pair was achieved.  In the next years new runs took place with the maximum energy of 200 GeV per nucleon pair. One of those runs was devoted to study deuteron--gold collisions which were analysed in order to get a reference point for more complicated gold--gold collisions. 

At present, the main activity in the field is connected with Large Hadron Collider (LHC) at CERN (Pb on Pb reactions at $\sqrt{s_{\rm NN}}$ = 2.76 TeV, first run in Nov.-Dec., 2010).  However, new experiments at lower energies (NA61 at CERN, STAR at BNL) are also very important, since this allows us to study the energy dependence of many characteristics of the particle production and analyse the systems at finite baryon chemical potential (see Marek Ga\'zdzicki's lecture).

\subsection{Theoretical tools}

In the relativistic heavy-ion collisions very large numbers (multiplicities) of particles are produced. For instance, in the central Au--Au collisions at RHIC, at the beam energy $\sqrt{s_{\rm NN}}$ = 200~GeV, the total charged particle multiplicity is about 5000. Hence, the number of produced particles exceeds the number of initial nucleons by a factor of~10. In this situation, different theoretical methods are used, which are appropriate for description of large macroscopic systems, e.g., thermodynamics, hydrodynamics, kinetic (transport) theory, field theory at finite temperature and density, non-equilibrium field theory, Monte-Carlo simulations. More importantly, we may also apply the fundamental theory of strong interactions.

In high-energy nuclear collisions a many-body system of strongly interacting particles is produced. The fundamental theory of strong interactions is Quantum Chromodynamics (QCD), the theory of quarks and gluons which are confined in hadrons, i.e., baryons and mesons. The discovery of asymptotic freedom in the strong interactions in 1973 by Gross, Politzer, and Wilczek \cite{Gross:1973id,Politzer:1973fx} allowed for making precise predictions of the results of many high-energy experiments in the framework of the perturbative quantum field theory  --- the asymptotic freedom is the property that the interaction between particles becomes weaker at shorter distances (see the School lectures by George Sterman). 
 
Probably the most striking feature of QCD is color confinement, which is the other side of the  asymptotic freedom. This is the phenomenon that color charged particles (such as quarks and gluons) cannot be isolated as separate objects. In other words, quarks and gluons cannot be directly observed. The physical concept of confinement may be illustrated by a string which is spanned between the quarks when we try to separate them. If the quarks are pulled apart too far, large energy is deposited in the string and it breaks into smaller pieces.  

\subsection{Quark-gluon plasma}

The main aim of the relativistic heavy-ion collisions is the observation of the two phase transitions predicted by QCD, i.e., the deconfinement and chiral phase transitions. At Earth conditions (i.e., at low energy densities) quarks and gluons are confined in hadrons. However, with increasing temperature (heating) and/or increasing baryon density (compression), a phase transition may occur to the state where ordinary hadrons do not exist any longer; quarks and gluons become the right degrees of freedom, and their motion is not confined to hadrons. 

This popular picture is based on the asymptotic freedom --- QGP is considered as an asymptotic state available at extremely high energies. Most likely, such a state has not been reached in the present experiments and, more importantly, it is very difficult to find an experimental evidence for its formation. On the other hand, we can accept a more pragmatic point of view and consider QGP as a new state of strongly interacting matter, whose properties can be inferred from experimental and theoretical investigations carried out at the currently available energies (with direct connections to QCD wherever it is possible).  The  present evidence suggests that the matter produced in heavy-ion collisions consists of quarks and gluons (due to the strong coupling these might not be elementary excitations in the system), it is locally well equilibrated, and characterised by the small (shear) viscosity to entropy density ratio. These striking experimental and theoretical findings suggest that QGP behaves more like a fluid than a gas \cite{Gyulassy:2004zy,Shuryak:2004cy}.

 In the limit of vanishing masses, the left-- and right--handed quarks become decoupled from each other and QCD becomes invariant under their interchange --- left- and right-handed quark currents are separately conserved, each state of the theory should have a degenerate partner of the opposite parity. On the other hand, we know that hadrons have well defined parity, and no such parity partners are observed. This paradox is resolved by the phenomenon of the spontaneous breakdown of chiral symmetry \cite{Nambu:1960xd,Nambu:1961tp}: the chiral symmetry of the interaction is broken by the true ground state of the theory. One expects that this symmetry is restored at high energies where quarks and gluons become the correct degrees of freedom. This is a very exciting subject but we are not going to follow it  any longer in these lectures. 

\section{Basic physics terms}
\label{sect:physters}

\subsection{Participants, spectators, and wounded nucleons} 

At high energies, simple geometric concepts are often used, for example, one distinguishes  participants from spectators --- if we assume that all nucleons propagate initially (i.e., before a collision) along parallel, straight line trajectories, then the nucleons which do not strike any other nucleons on their way are called spectators. Other nucleons which interact with each other are called participants. The participants which suffered at least one inelastic collision are called the wounded nucleons. A two-dimensional vector connecting centres of the colliding nuclei  is called the impact vector (its length is the impact parameter).  In particle and nuclear physics one introduces a coordinate system, where the spatial $z$-axis is parallel to the beam, and the impact vector ${\bf b}$ points in $x$-direction.  The axes $x$ and $z$ span the reaction plane of a given collision.

The simple picture outlined above is most convenient for theoretical investigations where we control the (initial) geometry of the collision process. On the other hand, the quantities such as the impact parameter or the reaction plane are not directly measured observables and we have to introduce them in a more sophisticated way. Before we do it, let us define the popular ways to parametrise the four-momenta of the produced particles. 

\subsection{Transverse mass and rapidity}

The component of a three-vector ${\bf A}$ parallel to $z$-axis is usually denoted by ${\bf A}_{\Vert }$, and the transverse component is 
${\bf A}_{\perp}={\bf A-A}_{\parallel}$. The transverse mass of a particle is defined as $m_{\perp}=\sqrt{m^2+{\bf p}_{\perp}^2}$,  where $m$ and ${\bf p}$ are the particle's mass and three-momentum~(the ``transverse'' quantities are also denoted by $T$, e.g., $m_T$ or $p_T$, the ``longitudinal'' quantities are then denoted by  $L$, e.g., $p_L$).

Since we deal with relativistic energies, it is useful to use the rapidity instead of the standard velocity
\begin{equation}
{\tt y}={\frac 12}\ln {\frac{(E+p_{\Vert })}{(E-p_{\Vert })}}=\hbox{arctanh}
\left( {\frac{p_{\Vert }}E}\right) =\hbox{arctanh}\left( {\tt v}_{\Vert} \right) . 
\label{rap}
\end{equation}
Here $E$ is the energy of a particle, $E=\sqrt{m^2+{\bf p}^2}$, and ${\tt v}_{\Vert}=p_{\Vert}/E$ is the longitudinal component of the velocity. 

Rapidity is additive under Lorentz boosts along the \mbox{$z$-axis}. This means that the difference $d{\tt y}$ as well as the rapidity
density $dN/d{\tt y}$ do not change under Lorentz boosts along the collision axis. The invariance under this type of transformation (corresponding to a constant $dN/d{\tt y}$) is shortly called the boost-invariance. 

Using the rapidity and the transverse mass, we can calculate the energy and the longitudinal momentum of a particle from the two equations: \mbox{$E=p^0=m_{\bot }\cosh {\tt y}$} and  $p_{\Vert }=m_{\bot }\sinh {\tt y}$. Experimentalists distinguish between rapidity and pseudorapidity. The latter is defined by the formula
\begin{equation}
\eta={\frac 12}\ln {\frac{(|{\bf p}|+p_{\Vert })}{(|{\bf p}|-p_{\Vert })}}=
\ln \left( \cot \frac \theta 2\right)= -\ln \left( \tan \frac \theta
2\right),
\label{psrap}
\end{equation}
where $\theta $ is the scattering angle. Pseudorapidity is easier to measure than rapidity (it is just a measure of the angle at which a particle has been emitted). To measure rapidity one has to identify the particle. Since at large energies $E \approx |{\bf p}|$ one is often tempted to assume that $dN/d{\tt y} \approx dN/d\eta$. In practice, this approximation is poor, especially in the region where rapidity is close to zero.

In theoretical calculations one usually uses the space-time rapidity
\begin{equation}
\eta_\parallel = {\frac 12}\ln \frac{t+z}{t-z}.
\label{strap}
\end{equation}
The space-time rapidity $\eta_\parallel$ together with the (longitudinal) proper time \mbox{$\tau=\sqrt{t^2-z^2}$} are used instead of the coordinates $t$ and $z$ to parametrise the interior of the light cone $t^2 =z^2 \, (t>0)$. One has to be careful to distinguish between rapidity, pseudorapidity, and space-time rapidity. In realistic calculations they are usually quite different. Only in very simple boost invariant models one may assume that these three quantities are equal.

\section{Hagedorn limiting temperature}
\label{sect:hag}

In 1960s the statistical bootstrap model (SBM) was introduced  \cite{Hagedorn:1965st,Hagedorn:1968ua} that was based on the observation that hadrons form bound and resonance states. This led to the concept of a possibly unlimited sequence of heavy resonance states, each being a constituent of a still heavier resonance. The number of such states in the mass interval $(m,m+dm)$ is denoted by $\rho(m) dm$, and the function  $\rho (m)$ is known as the SBM mass spectrum. The requirement that resonances are formed from other resonances in the self-consistent manner leads to the bootstrap condition for the mass spectrum $\rho (m)$. The solution of the bootstrap equation shows that the mass spectrum for large masses $m$ grows exponentially, as found by Hagedorn already in 1965 \cite{Hagedorn:1965st}. As a consequence, any thermodynamics employing this mass spectrum has a singular temperature $T_{\rm H}$ generated by the asymptotics $\rho (m)\sim \exp (m/T_{\rm H})$. At present $T_{\rm H}$ (Hagedorn temperature) is interpreted as the temperature where the phase transition from the hadron gas to the quark-gluon plasma occurs \cite{Cabibbo:1975ig}.

The subject of the limiting temperature is still an intriguing issue. More recent studies of the hadron mass spectrum have revealed that the Hagedorn temperatures of mesons and baryons are different \cite{Broniowski:2000bj,Broniowski:2004yh}.  The concepts that we are still missing some resonance states and they may be responsible for (fast) thermalization of the produced matter are widely discussed \cite{NoronhaHostler:2007jf}.

\section{Harmonic flows}
\label{sect:harmflow}

At present, the extraction of the reaction plane is one aspect of the very advanced  flow analysis of the collisions \cite{Ollitrault:1992bk,Heinz:2013th}. In this type of the investigations one represents the momentum distribution of the produced particles in the form
\begin{equation}
\frac{dN}{d{\tt y} d^2p_\perp} = \frac{dN}{2 \pi p_\perp dp_\perp d{\tt y}}
\left[1 + \sum\limits_{k=1}^\infty 2 v_k \cos\left( k (\phi_p - \Psi_k) \right)
\right], 
\label{flow-an}
\end{equation}
where $\Psi_k$ is the reference angle defined by the condition $\langle \sin(k \Psi_k)\rangle =0$, where the averaging is done over all particles in one event. Until very recently it has been common to assume $\Psi_k = \Psi_{\rm RP}$ for all $k$'s and to identify the angle $\Psi_{\rm RP}$ with the angle which specifies the position of the reaction plane.

Averaging of (\ref{flow-an}) over the azimuthal angle gives the transverse-momentum distribution. The coefficients $v_k$ characterize the momentum anisotropy. The coefficient $v_1$ is called the directed flow, whereas the coefficient $v_2$ is called the elliptic flow. In general, the coefficients $v_k$ are functions of rapidity and transverse momentum, 
$v_k = v_k({\tt y},p_\perp)$, and in this form often called the $k$th harmonic differential flow.
 
\subsection{Directed flow}

At low energies, the directed flow is manifested by the reflection of incoming particles by the first produced regions of highly compressed nuclear matter.  At the SPS energies the situation is already quite complex \cite{Stoecker:2004qu}. At positive rapidities the proton directed flow is positive, while the pion directed flow is negative. This suggests a different origin of $v_1$ of protons and pions. At the RHIC energies the directed flow of charged particles is negative, whereas the $v_1$ of the spectator neutrons is positive. This trend in the data suggests different behavior of the matter created in the central region and in the target/projectile fragmentation regions \cite{Bialas:2004su,Bozek:2010bi}. 

\subsection{Elliptic flow and thermalisation} 
\label{sect:v2}

In a non-central collision (characterised by a finite impact parameter~$b$), the particles are produced in an almond-like region in the transverse plane. This region can be characterised by the spatial anisotropy parameter $\epsilon$. The initial longitudinal momenta of the produced particles are very large (due to the initial impact). On the other hand, the typical initial transverse momenta are small and distributed isotropically. If the produced particles do not interact, their final transverse momenta should be also isotropic. Contrary, if the particles do interact, the spatial anisotropy is being transferred into the momentum anisotropy \cite{Ollitrault:1992bk}. This leads to non-zero values of the coefficient $v_2$ in Eq.~(\ref{flow-an}), which becomes proportional to $\epsilon$.

At RHIC energies, the measured values of the elliptic flow were explained first by the approaches based on the perfect-fluid hydrodynamics \cite{Huovinen:2001cy}. This led to the conclusion that the system produced in such collisions is characterised by the very small shear viscosity $\eta$ \cite{Gyulassy:2004zy,Shuryak:2004cy}. Further studies done within dissipative hydrodynamics set an upper limit on the ratio of the shear viscosity $\eta$ to the entropy density $s$, which is still very low \cite{Romatschke:2007mq,Heinz:2011kt}, $1 \leq 4\pi \eta/s \leq 3$ (smaller than for Helium at its critical temperature). One has to clarify that the shear viscosity itself  is not a good measure of the viscous effects as it is a dimensional quantity. A better measure is the ratio $\eta/s$ as it is a dimensionless observable (in the natural system of units). Therefore, the recent hydrodynamic studies aim at fixing $\eta/s$. 

Originally, the large values of $v_2$ were treated as the evidence of fast thermalisation of the produced matter. A frequently repeated statement was that $v_2$ could be generated only in the very early times, smaller than~1~fm/$c$. Model calculations show, however,  that the elliptic flow can be generated during the whole time evolution of the system \cite{Bialas:2007gn,Ryblewski:2010tn} (although the initial growth is the strongest). The realistic values of $v_2$ can be obtained with scenarios assuming initial free streaming of patrons or large anisotropy of the initial pressure \cite{Broniowski:2008qk,Ryblewski:2012rr}. This allows for delayed thermalisation  taking place at the times 1--2 fm/$c$, which is compatible with the results of different microscopic calculations (see the School lectures by Francois Gelis and Michael Strickland).

\subsection{Triangular flow}

Recent hydrodynamic calculations include event-by-event fluctuations in the distribution of the initial energy or entropy density  \cite{Gale:2012rq,Gale:2012in} (this effect follows from the fluctuations of the nucleon-nucleon collisions which are described below in the framework of the Glauber model). The fluctuations make this distribution less symmetric and this gives rise to both cosine and sine terms in the Fourier decomposition of the particle momenta in the azimuthal angle.  Equivalently, different reference angles $\Psi_k$ should be used for each value of $k$ in the decomposition (\ref{flow-an}). In event-by-event hydrodynamics, one performs the hydrodynamic calculation for a single event first, extracts $v_k$'s, and then the averaging over the events is done. This leads to non-trivial results for the odd harmonics such as, for example, $v_3$ which is called the triangular flow \cite{Alver:2010gr}. The combined measurements  of the elliptic and triangular flow enhance our ability to determine the value of the shear viscosity.

\section{Glauber model}
\label{sect:glauber}

Glauber model is used to determine the initial energy (or entropy) density of matter produced in heavy-ion collisions.~\footnote{For a recent review of the applications of the Glauber model to describe initial stages of relativistic heavy-ion collisions see, for example, Sec. III in Ref.~\cite{Florkowski:2010zz}. } With an additional assumption that the matter is thermalised, one can use the results obtained with the Glauber model as the input for hydrodynamic calculations which determine the subsequent space-time evolution of matter until the freeze-out point.

Originally, the Glauber model was applied only to elastic collisions. In this case a nucleon does not change its properties in the individual collisions, so all nucleon interactions can be well described by the same (elastic) cross section. Applying the Glauber model to inelastic collisions, we assume that after a single inelastic collision an excited nucleon-like object is created that interacts basically with the same inelastic cross section with other nucleons.

An alternative to the Glaber model calculations are theoretical studies related more directly to QCD. The most common approach of this type is the color--glass--condensate (CGC) theory \cite{McLerran:1993ni,McLerran:1993ka} which is based on the concept of gluon saturation \cite{GolecBiernat:1998js,GolecBiernat:1999qd} (see the lectures by Francois Gelis, Yuri Kovchegov and Larry McLerran).

\subsection{Nucleon-nucleon collisions}

At high energies, the inelastic nucleon-nucleon cross section $\sigma _{\rm {in}}$ gives the main contribution to the total cross section.  A certain subclass of the inelastic processes is the diffractive dissociation process. In this process a nucleon is only slightly excited and a small number of particles is produced. The diffractive processes contribute to about 10\% of all inelastic collisions. In non-diffractive inelastic nucleon-nucleon collisions a certain number of charged particles is produced. The average charged particle multiplicity ${\overline  N}_{\rm NN} $  is described by phenomenological formulas which give ${\overline  N}_{\rm NN} $ as a function of the energy $\sqrt{s}$ . Such formulas can be used in the comparisons done between heavy-ion and hadronic collisions.

Let us consider a nucleon-nucleon collision at a given energy $\sqrt{s}$ and at an impact parameter $b$.  In the eikonal approximation, we may introduce the probability of having a nucleon-nucleon inelastic collision
\begin{equation}
p\left( {\bf b}\right) =
\left(1- \left|e^{i \chi({\bf b})}\right|^2 \right) \equiv
t\left( {\bf b}\right) \sigma _{\rm {in}},
\label{pofb}
\end{equation}
where $\chi({\bf b})$ is the phase shift (times a factor of two). The function $t\left( {\bf b}\right)$, defined by (\ref{pofb}), is called the nucleon-nucleon  thickness function. The integral of $\,p\left( {\bf b} \right)$ over the whole range of the impact parameter should be normalized to $\sigma _{\rm {in}}$. Thus, the thickness function is normalized to unity.

\subsection{Nucleon-nucleus collisions}

The probability of finding a nucleon in the nucleus with the atomic mass number $A$ is the usual baryon density divided by the number of baryons in the nucleus. For large nuclei, one commonly uses the Woods-Saxon function  (our definition of $\rho_A(r)$ includes $A$ in the denominator, because we want to interpret $\rho_A(r)$ as the probability distribution)
\begin{equation}
\rho_A(r) =   \frac{\rho_0}{A} \,\left(1 + \exp \left[\frac{r-r_0}{a} \right]\right)^{-1},
\label{wsaxon}
\end{equation}
with the parameters: $r_0 = (1.12 A^{1/3}-0.86 A^{-1/3}) \, \hbox{fm}$, $a = 0.54 \, \hbox{fm}$, $\rho_0=0.17 \, \hbox{fm}^{-3}$. The parameter $\rho_0$ is the nuclear saturation density.

The nucleon-nucleus thickness function for the nucleus $A$ is obtained from a simple geometric consideration and the assumption that the nucleon positions in the nucleus $A$ are not changed during the collision process,
\begin{equation}
T_A\left({\bf b}\right) =
\int dz_A \int d^{\,2}s_A \, \rho_A({\bf s}_A,z_A)\, t({\bf s}_A-{\bf b}).  
\label{TA1}
\end{equation}
Here the transverse coordinates in the nucleus  $A$ are denoted by the vector ${\bf s}_A$ and
\begin{equation}
\rho_A({\bf s}_A,z_A) = \rho_A\left(\sqrt{{\bf s}_A^2+z_A^2}\,\right). 
\label{rhoA}
\end{equation}
Our definition of the Woods-Saxon distribution implies the normalization condition
\begin{equation}
\int d^{\,2}b \, T_A\left({\bf b}\right) = 1.  
\label{normTAofb}
\end{equation}

The quantity $T_A\left({\bf b}\right) \sigma_{\rm {in}}$ is the probability that a single nucleon-nucleon collision takes place in a nucleon-nucleus collision at the impact parameter ${\bf b}$. Treating all possible nucleon-nucleon collisions in the nucleon-nucleus collision as completely independent and characterized by the same cross section, we easily find the probability of having $n$ such collisions
\begin{equation}
P\left(n; A; {\bf b}\right) = 
\left( \begin{array}{c} A \\ n  \end{array} \right)
\left[ 1-T_{A}\left(  {\bf b}\right) \sigma
_{\rm {in}}\right] ^{A-n}\left[ T_{A}
\left(  {\bf b}\right) \sigma _{\rm {in}}\right] ^{n}.
\label{PAnb}
\end{equation}
The average number of binary nucleon-nucleon collisions may be calculated from the expression ${\overline n} \left(A;{\bf b}\right) =A\,T_{A}\left( {\bf b}\right) \,\sigma _{\rm {in}}$.

\subsection{Nucleus-nucleus collisions}

The thickness function for the nucleus-nucleus collision follows also from the geometric considerations which give in this case
\begin{equation}
T_{AB} \left({\bf b}\right) =
\int dz_A \int d^{\,2}s_A \, \rho_A({\bf s}_A,z_A) 
\int dz_B \int d^{\,2}s_B \, \rho_B({\bf s}_B,z_B) 
\,t({\bf b}+{\bf s}_B-{\bf s}_A), 
\label{TAB}
\end{equation}
with the corresponding normalization condition $\int d^{\,2}b \, T_{AB} \left({\bf b}\right) = 1$.

The quantity $T_{AB}\left({\bf b}\right) \sigma_{\rm {in}}$ is the averaged probability that a nucleon-nucleon collision takes place in a nucleus-nucleus collision characterized by the impact parameter ${\bf b}$.  In the limit $t({\bf b}) \rightarrow \delta^{(2)}({\bf b})$ we may write
\begin{eqnarray}
T_{AB} \left({\bf b}\right) 
&=& \int d^{\,2}s_A \, T_A({\bf s}_A)  
\, T_B({\bf s}_A-{\bf b}). \nonumber \\
\label{TAB1}
\end{eqnarray}

The nucleus-nucleus thickness function $T_{AB}\left({\bf b}\right)$ can be used to calculate the probability of having $n$ inelastic binary nucleon-nucleon collisions in a nucleus-nucleus collision at the impact parameter $\mathbf{b}$
\begin{equation}
P\left(n; AB; {\bf b}\right) = 
\left( \begin{array}{c} AB \\ n  \end{array} \right)
\left[ 1-T_{AB}\left(  {\bf b}\right) \sigma
_{\rm {in}}\right] ^{AB-n}\left[ T_{AB}
\left(  {\bf b}\right) \sigma _{\rm {in}}\right] ^{n}.
\label{PABnb}
\end{equation}
The average number of the collisions is ${\overline n} \left(AB;{\bf b}\right) = A B\,T_{AB}\left( {\bf b}\right) \,\sigma _{\rm {in}}$.

The total probability of an inelastic nuclear collision is the sum over $n$ from $n=1$ to $n=AB$
\begin{equation}
P_{\rm {in}}\left(AB;{\bf b} \right) =\sum_{n=1}^{AB}P\left(n; AB;{\bf b} \right) 
=1-\left[1-T_{AB}\left({\bf b}\right) \sigma_{\rm {in}}\right] ^{AB}.  
\label{PABb}
\end{equation}
In more realistic calculations, the positions of nucleons in the target and projectile nucleus are fixed, and the averaging should be done later. The probability of an inelastic collision for a fixed nucleon configuration equals
\begin{equation}
1 - \prod_{j=1}^A \prod_{i=1}^B 
\left[1 - t\left( {\bf b}+{\bf s}_{i}^{B}-{\bf s}_{j}^{A}\right) \sigma _{\rm {in}} \right].
\label{fixed-pos}
\end{equation}
The probability of an inelastic nuclear collision at the impact parameter $\mathbf{b}$ is then 
\begin{eqnarray}
\nonumber \\
P_{\rm {in}}\left(AB;{\bf b} \right) &=& \int d^2 s^A_1 T_A({\bf s}_{1}^{A}) \cdots d^2 s^A_A T_A({\bf s}_{A}^{A})
\int d^2 s^B_1 T_B({\bf s}_{1}^{B})\cdots d^2 s^B_B T_B({\bf s}_{B}^{B})
\nonumber \\
& & \times \left\{ 1 - \prod_{j=1}^A  \prod_{i=1}^B 
\left[1 - t\left( {\bf b}+{\bf s}_{i}^{B}-{\bf s}_{j}^{A}\right) \sigma _{\rm {in}} \right] \right\}.
\label{PABb-true} \\ \nonumber
\end{eqnarray}
The integration of (\ref{PABb-true}) over $b$ gives $\sigma_{\rm {in}}^{AB}$. Equations (\ref{PABb}) and (\ref{PABb-true}) differ from each other! The more accurate formula (\ref{PABb-true}) is much more complicated to handle and cannot be simply reduced to (\ref{PABb}). Only for nucleon-nucleus collisions the two methods are equivalent. Since there is no good analytic method to evaluate (\ref{PABb-true}) for large values of $A$ and $B$, one is most often satisfied with Eq. (\ref{PABb}) only. It is called the optical limit of the Glauber model. In order to have a more reliable distributions one does the Monte-Carlo calculations \cite{Broniowski:2007nz,Rybczynski:2013yba}.

\subsection{Wounded nucleons}

The Glauber model can be used also to calculate the number of the participants. To be more precise we distinguish between the participants which may interact elastically and the participants which interact only inelastically. The latter are called the wounded nucleons \cite{Bialas:1976ed}.  The number of nucleons in the nucleus $A$ is $A\int d^{2}s\,T_{A} ({\bf s})$. Probability, that the nucleon from $A$ at the position ${\bf s}$ collides one or more times with the nucleons in $B$ (in an $AB$ collision at the impact parameter ${\bf b}$) is
\begin{eqnarray}
\sum\limits_{n=1}^{B} P\left(n; B; {\bf b-s}\right) = 1-\left[ 1-\sigma
_{\rm {in}}T_{B}\left( {\bf b-s }\right) \right] ^{B}. \nonumber \\
\label{woundedb1a}
\end{eqnarray}
The number of wounded nucleons in $A$ can be obtained from the expression
\begin{eqnarray}
\overline{w}_A \left( A;B;{\bf b}\right)&=&
A\int d^{2}s\,T_{A}\left({\bf s}\right) \left( 1-\left[ 1-\sigma
_{\rm {in}}T_{B}\left( {\bf b-s}\right) \right] ^{B}\right).
\label{woundedb1b}
\end{eqnarray}
An analogous expression holds for the number of wounded nucleons in $B$. Since the number of wounded nucleons in the collision of $A$ and $B$ is the sum of the wounded nucleons in the nucleus $A$ and $B$, we obtain \cite{Bialas:1976ed} (after making the appropriate shifts in the integration over positions ${\bf s}$)
\begin{eqnarray}
\overline{w}\left( A;B;{\bf b}\right)&=&
A\int d^{2}s\,T_{A}\left( {\bf b}-{\bf s}\right) \left( 1-\left[ 1-\sigma
_{\rm {in}}T_{B}\left( {\bf s}\right) \right] ^{B}\right)  \nonumber \\
&& + \, B\int d^{2}s\,T_{B}\left( {\bf b}-%
{\bf s}\right) \left( 1-\left[ 1-\sigma _{\rm {in}}T_{A}\left( {\bf s}\right) %
\right] ^{A}\right). 
\label{woundedb}
\end{eqnarray}

\subsection{Nuclear modification factor}

A simple way to quantify the differences between the nucleus-nucleus collisions and the nucleon-nucleon collisions is to calculate the nuclear modification factor, 
\begin{eqnarray}
\nonumber \\
R_{AB}(p_\perp) &=& \frac{1}{ {\overline n}_{AB}}
\frac{d^2{\overline N}_{AB}}{ dp_\perp d\eta} /
\frac{1}{ \sigma^{pp}_{\rm tot}}
\frac{d\sigma^{pp}_{\rm incl} }{ dp_\perp d\eta}. 
\label{RAB} \\ \nonumber 
\end{eqnarray}
${\overline N}_{AB}$ -- average number of particles produced in the collisions of the nuclei $A$ and $B$, ${\overline n}_{AB}$ -- number of the binary nucleon-nucleon collisions obtained in the framework of the Glauber model. \medskip

The denominator of (\ref{RAB}) is the inclusive cross section for $pp$ collisions divided by the total cross section. This quantity is equal to the average number of particles produced in $pp$ collisions in the appropriate phase-space interval,
\begin{eqnarray}
\nonumber \\
\frac{dN_{pp} }{  dp_\perp d\eta} &=& \frac{1}{ \sigma^{pp}_{\rm tot}}
\frac{d\sigma^{pp}_{\rm incl} }{ dp_\perp d\eta}. 
\label{inclusivecs} \\ \nonumber 
\end{eqnarray}
If the collisions of the nuclei $A$ and $B$ are simple superpositions of the elementary $pp$ collisions, the scaling with the number of binary collisions should hold, and the nuclear modification factor is expected to be equal to unity. This is the reason why the nuclear modification factor  is so frequently used in direct searches for non-trivial dynamics.

\section{Hydrodynamic description of heavy-ion collisions}
\label{sect:hydro}

\subsection{Early statistical and hydrodynamic models}

The use of relativistic hydrodynamics to describe particle production in hadronic collisions has a long history which starts with the famous work of Landau in the early 1950s \cite{Landau:1953gs}, see also \cite{Khalatnikov:1954,Belenkij:1956cd}.  Landau's considerations were preceded, however, by a few approaches that used pure statistical and thermodynamic concepts in the analysis of hadronic collisions \cite{Koppe:1948,Fermi:1950jd,Fermi:1951zz,Pomeranchuk:1951ey}. Such approaches may be regarded as pre-hydrodynamic models and it is important to present them before we turn to the discussion of the genuine hydrodynamic models. 

\subsubsection{Fermi model}

Fermi assumed that when two relativistic nucleons collide, the deposited energy is released in a very small volume $V$, whose magnitude corresponds to the Lorentz contracted characteristic pion field volume $V_0$, i.e., $V = 2 m_{\rm N} V_0/\sqrt{s}$, where $V_0 = (4/3) \, \pi R_\pi^3$ with $R_\pi = 1/m_\pi$, and where $\sqrt{s}$ is the center-of-mass energy.  Subsequently, such a dense system decays into one of many accessible multiparticle states. The decay probability is calculated in the framework of the standard statistical physics.
        
The main reason for the introduction of the statistical concepts was the breakdown of the perturbation theory in description of strongly interacting systems. Clearly, the large values of the coupling constant prohibit the application of the perturbation theory. On the other hand, the large coupling is responsible for the phenomenon of multiparticle production, which is a very characteristic feature of strong interactions. 

The probability of the transition into a given state is proportional to the square of the matrix element and to the density of states.  In the statistical description, the matrix elements are treated as constants and the main effect comes from the phase space. Thus, the statistical approach represents a simple theoretical modeling of collisions which may be regarded as the complementary approach to the perturbation schemes which typically break down at a certain scale. The main heuristic argument for the justification of the use of the statistical approach is that the role of the phase space naturally grows with the increasing energy of the collisions. 
    
According to Fermi, the probability for the formation of the state with $n$ particles is proportional to the factor
\begin{equation}
S(n) = \left[ \frac{V}{(2\pi)^3} \right]^{(n-1)} \frac{dQ(W)}{dW}.
\label{fermi-th-1}
\end{equation}
Here $W$ is the total energy of the colliding system, $dQ/dW$ is the number of states per unit energy, and $V$ is the interaction volume. The power $n-1$ arises from the fact that  the momenta of only $n-1$ particles are independent. 

In the last chapter of his seminal paper from 1950, Fermi considers the collisions at extremely high energies. He argues that in this case a detailed statistical considerations may be replaced by the simple thermodynamic arguments. Assuming that the matter is thermalized one can calculate the temperature of the produced hadronic system from the thermodynamic relations valid for massless particles.  Fermi took into account only the production of pions, nucleons and antinucleons, and used the formula
\begin{equation}
\left( \varepsilon_\pi + \varepsilon_{N+{\bar N}} \right) \, V \,
= \frac{\pi^2 V T^4}{3} = W.
\label{st-th-1} 
\end{equation}
Using the expression for the Lorentz contracted volume we rewrite (\ref{st-th-1}) in the following form
\begin{equation}
T^4 = \frac{3}{2\pi^2} \frac{W^2}{V_0 m_{\rm N}} = \frac{9}{8\pi^3} \frac{W^2 m_\pi^3}{m_{\rm N}}.
\label{st-th-2} 
\end{equation}
This equation may be used to calculate the abundances of the produced pions, nucleons and antinucleons from the thermodynamic relations giving the particle densities in terms of the temperature. It is interesting to note that Fermi's idea forms the ground for present thermal model analyses discussed below in Sec.~\ref{sect:chem}.

\subsubsection{Landau model}

In the Landau model \cite{Landau:1953gs} an expansion of matter before the hadron decoupling is included.  The idea of modification of the Fermi approach was indicated by Pomeranchuk \cite{Pomeranchuk:1951ey}. He argued that the particles in the system should interact until the average distance between them becomes larger than the typical interaction distance.  Landau proposed his hydrodynamic approach to describe proton-proton collisions. Following Fermi, he assumed that the two colliding protons released their energy in the volume corresponding to the Lorentz-contracted size of a proton.  Under the influence of the longitudinal gradient the system starts expanding. The transverse gradient is also present but initially the gradient in the longitudinal direction is much larger and the early expansion may be regarded as one-dimensional. 

\subsubsection{Bjorken model}

In the Landau model the initial conditions are specified for a given laboratory time in the center-of-mass frame, when the matter is highly compressed and at rest. Landau's description does not include one aspect of high-energy processes -- fast particles are produced later and further away from the collision center than the slow particles. It is possible to include this effect by imposing special initial conditions. This idea was proposed and developed by Bjorken \cite{Bjorken:1982qr}. 

The Bjorken hydrodynamic model \cite{Bjorken:1982qr} is based on the assumption that the rapidity distribution $dN_{\rm ch}/d{\tt y}$ is constant in the mid-rapidity region. This fact means that the central region is boost invariant. In this case the longitudinal flow has the form $v_z=z/t$ and all thermodynamic quantities characterizing the central region depend only on the longitudinal proper time \mbox{$\tau =\sqrt{t^2-z^2}$} and the transverse coordinates $x$ and $y$.

The main success of the Bjorken model is that it allows for simple and realistic estimates of the initial energy density available at the early stages of heavy-ion collisions. Such estimates always indicate that the produced matter has the energy density much larger than the typical energy density characterising the phase transition from the hadron gas to the quark-gluon plasma. Thus, using Bjorken's simple estimates we expect a formation of a new state of matter in heavy-ion collisions at the relativistic energies. 

\subsubsection{Towards modern hydro approaches}

Modern hydrodynamic calculations follow general concepts introduced in the Landau and Bjorken models. However, they differ from the Landau original description in the way how they treat the initial conditions. They also go beyond the simple Bjorken approach by including transverse expansion \cite{Baym:1983sr} and, eventually, by breaking the boost-invariance.  Additionally, the recent hydrodynamic codes use modern equations of state inspired by the lattice simulations of QCD and advanced hadron-gas calculations. In addition, the hydrodynamic simulations are performed on the event-by-event basis \cite{Gale:2012rq,Gale:2012in} .

In the Landau and Bjorken models, the thermalized matter was a gas of ultra-relativistic particles, mainly pions, satisfying the extreme relativistic equation of state $P = \varepsilon/3$. At present, more accurate equations of state for hot and dense matter are known (for the perturbative QCD calculation see \cite{Haque:2013sja,Haque:2014rua} and for the lattice QCD calculations see \cite{Borsanyi:2010cj,Bhattacharya:2014ara} and Christian Hoelbling's lecture). Expecting the phase transition to the quark-gluon plasma, one can use the plasma equation of state including the phase transition back to ordinary hadronic matter, see for example Refs.~\cite{Chojnacki:2007jc,Broniowski:2008vp}. In this way the phase transition is incorporated in the hydrodynamic frameworks in a very elegant and thermodynamically consistent way. 

It is interesting to stress that the use of the appropriate QCD equation of state was crucial for the correct description of the HBT radii (see Adam Kisiel's lecture). Moreover, determination of the real character of the phase transition may be helpful to perform realistic calculations within the cosmological models based on the Friedman equation \cite{Florkowski:2010mc}.

\subsection{Initial conditions}

For boost-invariant systems with vanishing baryon chemical potential one usually assumes that either the initial entropy density, $\sigma_{\rm i} ({\bf x}_\perp) = \sigma(\tau_{\rm i},{\bf x}_\perp)$, or the initial energy density, $\varepsilon_{\rm i} ({\bf x}_\perp)=\varepsilon(\tau_{\rm i},{\bf x}_\perp)$, are directly related to the density of sources of particle production, $\rho_{\rm sr} ({\bf x}_\perp)$. \medskip

The sources considered in this context are wounded nucleons or binary collisions. The symmetry with respect to the Lorentz boosts along the collision axis means that it is sufficient to consider all these quantities in the plane $z=0$. In general, a mixed model is used, with a linear combination of the wounded-nucleon density ${\overline w}\left( {\bf x}_\perp \right)$ and the density of binary collisions ${\overline n} \left( {\bf x}_\perp \right)$. This leads to the following expression for the initial entropy \cite{Wang:2000bf,Kharzeev:2000ph}:
\begin{equation}
\sigma_{\rm i} ({\bf x}_\perp) \;\propto\; 
\rho_{\rm sr} ({\bf x}_\perp) = 
\frac{1-\kappa}{2}\, {\overline w}\left( {\bf x}_\perp \right) 
+ \kappa\, {\overline n} \left( {\bf x}_\perp \right),
\label{eqn:initial_entropy}
\end{equation}
where $\kappa$ is a fit parameter. The initial longitudinal profiles are less known. One usually uses gaussian parametrisation of the entropy density in space-time rapidity (with a possible flat insert in the middle). The width of this distribution is chosen in such a way as to reproduce the measured rapidity distribution.

\subsection{Hydrodynamics of perfect fluid}

The {\it perfect fluid} is defined formally by the form of its energy-momentum tensor, namely 
\begin{equation}
T_{\rm eq}^{\mu \nu }=\ ( \varepsilon +P)u^\mu u^\nu -Pg^{\mu \nu },  
\label{ten}
\end{equation}
where $g^{\mu \nu}$ is the metric tensor with $g^{00}=1$, $\varepsilon$ is the energy density, $P$ is the pressure, and $u^\mu$ is the four-velocity of the fluid element.

The form (\ref{ten}) follows from the assumption of local thermal equilibrium. Equations of motion of the perfect fluid are then obtained from the conservation laws 
\begin{equation}
\partial _\mu T_{\rm eq}^{\mu \nu }=0.  
\label{divten}
\end{equation}
In (\ref{divten}) we have four equations. On the other hand, we search for five unknown functions: three independent components of the fluid four-velocity, the energy density, and pressure. The system of equations becomes closed if (\ref{divten}) is supplemented with the equation of state, for example, in the form $P=P(\varepsilon)$. Using (\ref{divten}) one may check that the entropy is conserved, hence the flow is adiabatic.  If the fluid has non-zero baryon density, one should add the baryon number conservation law to (\ref{divten}) as an extra equation.

\subsection{Hydrodynamics of viscous fluid}

In viscous hydrodynamics, the energy-momentum tensor has the form
\begin{eqnarray}
T^{\mu\nu} = T_{\rm eq}^{\mu\nu} + \Pi^{\mu\nu},
\label{tendis}
\end{eqnarray}
where $T_{\rm eq}^{\mu\nu}$ is the perfect-fluid part given by (\ref{ten}) and $\Pi^{\mu\nu}$ describes dissipation
\begin{eqnarray}
\Pi^{\mu\nu} = \pi^{\mu\nu} + \Pi \Delta^{\mu \nu}, \quad \Delta^{\mu\nu} = g^{\mu \nu} - u^\mu u^\nu.
\label{Pimunu}
\end{eqnarray}
Here $\pi^{\mu\nu}$ is the shear tensor and $\Pi$ describes the viscous bulk pressure. The equations of hydrodynamics follow from the conservation laws for energy and momentum, and from the requirement that the entropy production is positive. These conditions determine the form of equations to be satisfied by the dissipative terms   $\pi^{\mu\nu}$ and $\Pi$. 

From the formal point of view, the inclusion of the dissipative terms in (\ref{tendis}) follows from the gradient expansion around the local equilibrium. In the first order in gradients one finds the Navier-Stokes expressions
\begin{equation}
\pi^{\mu\nu} = \eta \nabla^{<\mu} u^{\nu>}, \quad \Pi = \zeta \partial_\alpha u^\alpha,
\label{ez}
\end{equation}
where the angle brackets project out the traceless symmetric part (the symmetric part is denoted by round brackets)
\begin{equation}
\nabla^{<\mu} u^{\nu>} = 2 \nabla^{(\mu} u^{\nu)} - \frac{2}{3} \Delta^{\mu\nu} \nabla_\alpha u^\alpha, \quad
\nabla^\alpha = \Delta^{\alpha \beta} \partial_\beta.
\end{equation}
The quantities $\eta$ and $\zeta$ in (\ref{ez}) are the shear and bulk viscosity, respectively. Unfortunately, the relativistic fluid dynamics based on the Navier-Stokes prescription suffers from problems connected with the acausal  transmission of signals. This is why the second-order theory had been developed by Israel and Stewart \cite{Israel:1979wp}. Within the second-order theory, the shear tensor  $\pi^{\mu\nu}$ and the bulk pressure $\Pi$ satisfy non-trivial dynamic equations. They are not any longer expressed by simple formulas such as (\ref{ez}). Moreover, the second-order theory requires that higher-order kinetic coefficients should be introduced.

At the moment, the formalism developed by Israel and Stewart is the most popular version of the dissipative hydrodynamics used to describe heavy-ion collisions. Usually only the shear viscosity is included in such calculations. There are, however, suggestions that the bulk viscosity may also play an important role \cite{Noronha-Hostler:2013gga}. More importantly, the second-order formalism may lead to unphysical behaviour at the early stages of the collisions or at the edges  of the produced system. Such issues are discussed in the lectures by Michael Strickland \cite{Strickland:2014pga} in the context of a new formulation of dissipative fluid dynamics (anisotropic hydrodynamics \cite{Strickland:2014pga,Florkowski:2010cf,Martinez:2010sc}).

\section{Freeze-out}
\label{sect:fout}

\subsection{Kinetic freeze-out}

The thermal or kinetic freeze-out is the stage in the evolution of matter when the hadrons practically stop to interact. In other words, the thermal freeze-out is a transition from a strongly coupled system (very likely evolving from one local equilibrium state to another) to a weakly coupled one (consisting of essentially free-streaming particles). 

It is triggered by the expansion of matter, which causes a rapid growth of the mean free path, $\lambda_{\rm mfp}$, of particles. The thermal freeze-out happens when the timescale connected with the collisions, $\tau_{\rm coll} \sim \lambda_{\rm mfp}$, becomes larger than the expansion timescale, $\tau_{\rm exp}$. In this case the particles depart from each other so fast that the collision processes become ineffective. We may formulate this condition as the inequality \cite{Hung:1997du}
\begin{equation}
\tau_{\rm coll} \geq \tau_{\rm exp}. 
\label{freezeoutcrit0}
\end{equation}

The magnitude of the collision time is determined by the product of the  average cross section and the particle density,
\begin{eqnarray}
\tau_{\rm coll} \sim \frac{1}{\sigma \, n}, 
\label{taucoll} 
\end{eqnarray}
whereas the magnitude of the expansion time is characterized by the divergence of the four-velocity field, $u^\mu$, describing the hydrodynamic flow of matter,
\begin{eqnarray}
\tau_{\rm exp} \sim \frac{1}{ \partial_\mu u^\mu}.
\label{tauexp} 
\end{eqnarray}
Very often a simplified criterion is assumed which says that the thermal freeze-out happens at the time when the mean free path of hadrons is of the same order as the size of the system. 

\subsection{Chemical freeze-out and thermal models}
\label{sect:chem}

Chemical freeze-out defines the stage where the hadron abundances are fixed --- it should precede the kinetic
freeze-out. The concept of the chemical freeze-out  is used in thermal models of particle production. 
In such models one assumes that a gas of  stable hadrons and resonances is formed (at the chemical 
freeze-out). The final (measured) multiplicities of hadrons consist of primary particles, 
present in the hot fireball, and of secondary particles coming from the decays of resonances.

There exist several versions of the thermal approach in the literature, for example, see \cite{Koch:1985hk,Cleymans:1992zc,Gazdzicki:1998vd,
Becattini:2000jw,BraunMunzinger:2001ip,Florkowski:2001fp,Torrieri:2004zz,Petran:2013lja,Stachel:2013zma}. The most popular is the
grand canonical version,  where one assumes that all hadron species are formed in local thermal
and chemical equilibrium. Other versions of the thermal approach assume chemical non-equilibrium 
in the strange sector of particles, or chemical non-equilibrium in both the strange and non-strange sectors~\footnote{The results of different versions of the thermal model applied at the LHC energy have been recently reviewed in \cite{Floris:2014pta}}. 

There is also a possibility that the chemical and thermal freeze-out coincide. Such a single freeze-out model was
constructed in Ref.~\cite{Broniowski:2001we} and was successfully used to describe the RHIC soft hadronic observables.
Its Monte-Carlo version is defined in Refs.~\cite{Kisiel:2005hn,Chojnacki:2011hb}.

The recent LHC data on heavy-ion collisions show that the predictions of two popular versions of the statistical
model (the chemical equilibrium model and the strangeness non-equilibrium model) give too large values for the kaon to pion
ratio, $(K^++K^-)/(\pi^++\pi^-)$, and, especially, for the ratio of protons to pions $(p+\bar{p})/(\pi^++\pi^-)$
\cite{Abelev:2012wca,Abelev:2013vea}. The recent fit \cite{Stachel:2013zma} gives almost three standard deviations 
higher values for protons and anti-protons compared to the LHC data. Besides the problems with thermal interpretation of the hadron
abundances at the LHC, one encounters also the problems with the hydrodynamic interpretation of the transverse-momentum spectra of
pions, kaons and protons. It turns out   that one can connect the proton puzzle with the anomalous behavior of the pion $p_T$ spectra and solve the two problems within the chemical non-equilibrium version of the  single freeze-out model \cite{Begun:2013nga,Begun:2014rsa}.

In the hydrodynamic calculations the freeze-out process is modelled in two alternative ways: either one uses
the concept of a single freeze-out and assumes that the hadrons are completely decoupled on a specific freeze-out
hypersurface or one switches from the hydrodynamic description to the hadronic cascade model which relies on the kinetic theory \cite{Petersen:2008dd}.

\section{Hanbury--Brown-Twiss interferometry}
\label{sect:hbt}

The fundamental object in  the HBT interferometry is the two-particle correlation function $C({\bf p}_1,{\bf p}_2)$, measured for pairs of identical particles such as $\pi ^{+}\pi ^{+}$, $\pi ^{-}\pi ^{-}$, or $K^{+}K^{+}$. In general, it is defined by the expression 
\begin{eqnarray}
C({\bf p}_1,{\bf p}_2) &=& \frac{ {\cal P}_2({\bf p}_1,{\bf p}_2)}
{{\cal P}_1({\bf p}_1) {\cal P}_1({\bf p}_2)}, 
\label{Cdef}  
\end{eqnarray}
where ${\cal P}_1({\bf p})$ is the invariant inclusive one-particle distribution function in the space of rapidity and transverse-momentum, 
\begin{eqnarray}
{\cal P}_1({\bf p}) &=& E_p \frac{dN}{d^3p} = \frac{dN}{d{\tt y}d^2p_\perp},
\label{P1}   
\end{eqnarray}
and ${\cal P}_2({\bf p}_1,{\bf p}_2)$ is the analogous two-particle distribution
\begin{eqnarray}
{\cal P}_2({\bf p}_1,{\bf p}_2) &=&  E_{p_1} E_{p_2} 
{dN \over d^3p_1 d^3p_2} = \frac{dN}{d{\tt y}_1 d^2p_{1 \perp} d{\tt y}_2 d^2p_{2 \perp}}.
\label{P2}   
\end{eqnarray}
Equations (\ref{P1}) and (\ref{P2}) imply that the correlation function (\ref{Cdef}) transforms like a Lorentz scalar. In (\ref{Cdef}) we may use the average momentum 
\begin{equation}
{\bf k} = \frac{1}{2}\,\left({\bf p}_1+{\bf p}_2\right), 
\label{bfk}
\end{equation}
and the difference of the two momenta
\begin{equation}
{\bf q} ={\bf p}_1-{\bf p}_2. 
\label{bfq}
\end{equation} 

In the analyses of the correlation functions one uses commonly the out-side-long coordinate system.  First, by making the Lorentz boost along the collision axis we may set ${\bf k}_\parallel =0$. In this way we change to the special frame that is called the longitudinally comoving system (LCMS).  In this frame, the direction of the beam is called the long direction. The direction of the three-momentum of the pair is the out direction, and the third orthogonal direction is called the side direction. The measured correlation functions are usually fitted with the gaussians of the following form~\footnote{Strictly speaking, the parametrisation (\ref{BPradii}) is suitable for boost-invariant and azimuthally symmetric systems. In more general cases one should use more complex formulas. }
\begin{equation}
C(k_\perp,{\bf q}) = 1 +  \lambda \exp
\left[ -R^2_{\rm long}(k_\perp) q^2_{\rm long} 
 -R^2_{\rm out}(k_\perp) q^2_{\rm out} 
-R^2_{\rm side}(k_\perp) q^2_{\rm side} \right]. 
\label{BPradii} 
\end{equation}
The parameters $R_{\rm side}$, $R_{\rm out}$, and $R_{\rm long}$ are called the HBT radii. The measurements of the HBT radii as functions of the mean transverse momentum of the pion pairs gives us information about the space-time sizes of the system at the kinetic freeze-out (for more details see the lecture by Adam Kisiel).

\section{Conclusions}
\label{sect:concl}

Successful applications of relativistic hydrodynamics in description of ultra-relativistic heavy-ion collisions allowed us to establish a uniform picture of these complicated processes. In some sense we are in a fortunate situation that such complex systems and processes may be described within a concise and well-defined framework. 

The hydrodynamic approach, combined with the modeling of the initial state by the Glauber model or the color glass condensate on one side, and supplemented by the kinetic simulations of the freeze-out process on the other side, forms the foundation of an approach that may be regarded as the standard model of ultra-relativistic heavy-ion collisions. Nevertheless, many details of this picture should be improved and surprises may wait for us just around the corner.


This work was supported in part by the Polish National Science Center with Decision No. DEC-2012/06/A/ST2/00390.




\newpage
\bibliographystyle{utphys}
\bibliography{wf}

\providecommand{\href}[2]{#2}\begingroup\raggedright\begin{thebibliography}{10}

\bibitem{Csernai:1994xw}
L.~Csernai, ``{Introduction to relativistic heavy ion collisions},''
{\em (John Wiley \& Sons, Chichester, 1994)} .

\bibitem{Wong:1995jf}
C.~Wong, ``{Introduction to high-energy heavy ion collisions},''
{\em (World Scientific, Singapore, 2004)} .

\bibitem{Letessier:2002gp}
J.~Letessier and J.~Rafelski, ``{Hadrons and quark - gluon plasma},''
{\em Camb.Monogr.Part.Phys.Nucl.Phys.Cosmol.} {\bfseries 18} (2002) 1--397.

\bibitem{Yagi:2005yb}
K.~Yagi, T.~Hatsuda, and Y.~Miake, ``{Quark-gluon plasma: From big bang to
  little bang},''
{\em Camb.Monogr.Part.Phys.Nucl.Phys.Cosmol.} {\bfseries 23} (2005) 1--446.

\bibitem{Florkowski:2010zz}
W.~Florkowski, ``{Phenomenology of Ultra-Relativistic Heavy-Ion Collisions},''
{\em (World Scientific, Singapore, 2010)} .

\bibitem{Friman:2011zz}
B.~Friman, C.~Hohne, J.~Knoll, S.~Leupold, J.~Randrup, {\em et~al.}, ``{The CBM
  physics book: Compressed baryonic matter in laboratory experiments},''
\href{http://dx.doi.org/10.1007/978-3-642-13293-3}{{\em Lect.Notes Phys.}
  {\bfseries 814} (2011) 1--980}.

\bibitem{Shuryak:1978ij}
E.~V. Shuryak, ``{Quark-gluon plasma and hadronic production of leptons,
  photons and pions},''
\href{http://dx.doi.org/10.1016/0370-2693(78)90370-2}{{\em Phys. Lett.}
  {\bfseries B78} (1978) 150}.

\bibitem{Shuryak:1980tp}
E.~V. Shuryak, ``{Quantum chromodynamics and the theory of superdense
  matter},''
\href{http://dx.doi.org/10.1016/0370-1573(80)90105-2}{{\em Phys. Rept.}
  {\bfseries 61} (1980) 71--158}.

\bibitem{Gross:1973id}
D.~J. Gross and F.~Wilczek, ``{Ultraviolet behavior of non-abelian gauge
  theories},''
\href{http://dx.doi.org/10.1103/PhysRevLett.30.1343}{{\em Phys. Rev. Lett.}
  {\bfseries 30} (1973) 1343--1346}.

\bibitem{Politzer:1973fx}
H.~D. Politzer, ``{Reliable perturbative results for strong interactions?},''
\href{http://dx.doi.org/10.1103/PhysRevLett.30.1346}{{\em Phys. Rev. Lett.}
  {\bfseries 30} (1973) 1346--1349}.

\bibitem{Gyulassy:2004zy}
M.~Gyulassy and L.~McLerran, ``{New forms of QCD matter discovered at RHIC},''
  \href{http://dx.doi.org/10.1016/j.nuclphysa.2004.10.034}{{\em Nucl.Phys.}
  {\bfseries A750} (2005) 30--63},
\href{http://arxiv.org/abs/nucl-th/0405013}{{\ttfamily arXiv:nucl-th/0405013
  [nucl-th]}}.

\bibitem{Shuryak:2004cy}
E.~V. Shuryak, ``{What RHIC experiments and theory tell us about properties of
  quark-gluon plasma?},'' {\em Nucl. Phys.} {\bfseries A750} (2005) 64--83.

\bibitem{Nambu:1960xd}
Y.~Nambu, ``{Axial vector current conservation in weak interactions},''
\href{http://dx.doi.org/10.1103/PhysRevLett.4.380}{{\em Phys. Rev. Lett.}
  {\bfseries 4} (1960) 380--382}.

\bibitem{Nambu:1961tp}
Y.~Nambu and G.~Jona-Lasinio, ``{Dynamical model of elementary particles based
  on an analogy with superconductivity. I},''
\href{http://dx.doi.org/10.1103/PhysRev.122.345}{{\em Phys. Rev.} {\bfseries
  122} (1961) 345--358}.

\bibitem{Hagedorn:1965st}
R.~Hagedorn, ``Statistical thermodynamics of strong interactions at high-
  energies,''
{\em Nuovo Cim. Suppl.} {\bfseries 3} (1965) 147--186.

\bibitem{Hagedorn:1968ua}
R.~Hagedorn and J.~Ranft, ``{Statistical thermodynamics of strong interactions
  at high-energies. 2. Momentum spectra of particles produced in $p p$
  collisions},''
{\em Nuovo Cim. Suppl.} {\bfseries 6} (1968) 169--354.

\bibitem{Cabibbo:1975ig}
N.~Cabibbo and G.~Parisi, ``Exponential hadronic spectrum and quark
  liberation,''
{\em Phys. Lett.} {\bfseries B59} (1975) 67.

\bibitem{Broniowski:2000bj}
W.~Broniowski and W.~Florkowski, ``{Different Hagedorn temperatures for mesons
  and baryons},'' {\em Phys. Lett.} {\bfseries B490} (2000) 223--227.

\bibitem{Broniowski:2004yh}
W.~Broniowski, W.~Florkowski, and L.~Y. Glozman, ``{Update of the Hagedorn mass
  spectrum},'' {\em Phys. Rev.} {\bfseries D70} (2004) 117503.

\bibitem{NoronhaHostler:2007jf}
J.~Noronha-Hostler, C.~Greiner, and I.~A. Shovkovy, ``{Fast equilibration of
  hadrons in an expanding fireball},'' {\em Phys. Rev. Lett.} {\bfseries 100}
  (2008) 252301.

\bibitem{Ollitrault:1992bk}
J.-Y. Ollitrault, ``{Anisotropy as a signature of transverse collective
  flow},''
\href{http://dx.doi.org/10.1103/PhysRevD.46.229}{{\em Phys.Rev.} {\bfseries
  D46} (1992) 229--245}.

\bibitem{Heinz:2013th}
U.~Heinz and R.~Snellings, ``{Collective flow and viscosity in relativistic
  heavy-ion collisions},''
  \href{http://dx.doi.org/10.1146/annurev-nucl-102212-170540}{{\em
  Ann.Rev.Nucl.Part.Sci.} {\bfseries 63} (2013) 123--151},
\href{http://arxiv.org/abs/1301.2826}{{\ttfamily arXiv:1301.2826 [nucl-th]}}.

\bibitem{Stoecker:2004qu}
H.~Stoecker, ``{Collective flow signals the quark gluon plasma},''
  \href{http://dx.doi.org/10.1016/j.nuclphysa.2004.12.074}{{\em Nucl.Phys.}
  {\bfseries A750} (2005) 121--147},
\href{http://arxiv.org/abs/nucl-th/0406018}{{\ttfamily arXiv:nucl-th/0406018
  [nucl-th]}}.

\bibitem{Bialas:2004su}
A.~Bialas and W.~Czyz, ``{Wounded nucleon model and Deuteron-Gold collisions at
  RHIC},'' {\em Acta Phys.Polon.} {\bfseries B36} (2005) 905--918,
\href{http://arxiv.org/abs/hep-ph/0410265}{{\ttfamily arXiv:hep-ph/0410265
  [hep-ph]}}.

\bibitem{Bozek:2010bi}
P.~Bozek and I.~Wyskiel, ``{Directed flow in ultrarelativistic heavy-ion
  collisions},'' \href{http://dx.doi.org/10.1103/PhysRevC.81.054902}{{\em
  Phys.Rev.} {\bfseries C81} (2010) 054902},
\href{http://arxiv.org/abs/1002.4999}{{\ttfamily arXiv:1002.4999 [nucl-th]}}.

\bibitem{Huovinen:2001cy}
P.~Huovinen, P.~Kolb, U.~W. Heinz, P.~Ruuskanen, and S.~Voloshin, ``{Radial and
  elliptic flow at RHIC: Further predictions},''
  \href{http://dx.doi.org/10.1016/S0370-2693(01)00219-2}{{\em Phys.Lett.}
  {\bfseries B503} (2001) 58--64},
\href{http://arxiv.org/abs/hep-ph/0101136}{{\ttfamily arXiv:hep-ph/0101136
  [hep-ph]}}.

\bibitem{Romatschke:2007mq}
P.~Romatschke and U.~Romatschke, ``{Viscosity Information from Relativistic
  Nuclear Collisions: How Perfect is the Fluid Observed at RHIC?},''
  \href{http://dx.doi.org/10.1103/PhysRevLett.99.172301}{{\em Phys.Rev.Lett.}
  {\bfseries 99} (2007) 172301},
\href{http://arxiv.org/abs/0706.1522}{{\ttfamily arXiv:0706.1522 [nucl-th]}}.

\bibitem{Heinz:2011kt}
U.~Heinz, C.~Shen, and H.~Song, ``{The viscosity of quark-gluon plasma at RHIC
  and the LHC},'' \href{http://dx.doi.org/10.1063/1.3700674}{{\em AIP
  Conf.Proc.} {\bfseries 1441} (2012) 766--770},
\href{http://arxiv.org/abs/1108.5323}{{\ttfamily arXiv:1108.5323 [nucl-th]}}.

\bibitem{Bialas:2007gn}
A.~Bialas, M.~Chojnacki, and W.~Florkowski, ``{Early evolution of transversally
  thermalized partons},''
  \href{http://dx.doi.org/10.1016/j.physletb.2008.02.042}{{\em Phys.Lett.}
  {\bfseries B661} (2008) 325--329},
\href{http://arxiv.org/abs/0708.1076}{{\ttfamily arXiv:0708.1076 [nucl-th]}}.

\bibitem{Ryblewski:2010tn}
R.~Ryblewski and W.~Florkowski, ``{Early anisotropic hydrodynamics and the RHIC
  early-thermalization and HBT puzzles},''
  \href{http://dx.doi.org/10.1103/PhysRevC.82.024903}{{\em Phys.Rev.}
  {\bfseries C82} (2010) 024903},
\href{http://arxiv.org/abs/1004.1594}{{\ttfamily arXiv:1004.1594 [nucl-th]}}.

\bibitem{Broniowski:2008qk}
W.~Broniowski, W.~Florkowski, M.~Chojnacki, and A.~Kisiel, ``{Free-streaming
  approximation in early dynamics of relativistic heavy-ion collisions},''
  \href{http://dx.doi.org/10.1103/PhysRevC.80.034902}{{\em Phys.Rev.}
  {\bfseries C80} (2009) 034902},
\href{http://arxiv.org/abs/0812.3393}{{\ttfamily arXiv:0812.3393 [nucl-th]}}.

\bibitem{Ryblewski:2012rr}
R.~Ryblewski and W.~Florkowski, ``{Highly-anisotropic hydrodynamics in 3+1
  space-time dimensions},''
  \href{http://dx.doi.org/10.1103/PhysRevC.85.064901}{{\em Phys.Rev.}
  {\bfseries C85} (2012) 064901},
\href{http://arxiv.org/abs/1204.2624}{{\ttfamily arXiv:1204.2624 [nucl-th]}}.

\bibitem{Gale:2012rq}
C.~Gale, S.~Jeon, B.~Schenke, P.~Tribedy, and R.~Venugopalan, ``{Event-by-event
  anisotropic flow in heavy-ion collisions from combined Yang-Mills and viscous
  fluid dynamics},''
  \href{http://dx.doi.org/10.1103/PhysRevLett.110.012302}{{\em Phys.Rev.Lett.}
  {\bfseries 110} (2013) 012302},
\href{http://arxiv.org/abs/1209.6330}{{\ttfamily arXiv:1209.6330 [nucl-th]}}.

\bibitem{Gale:2012in}
C.~Gale, S.~Jeon, B.~Schenke, P.~Tribedy, and R.~Venugopalan, ``{Initial state
  fluctuations and higher harmonic flow in heavy-ion collisions},''
  \href{http://dx.doi.org/10.1016/j.nuclphysa.2013.02.037}{{\em Nucl.Phys.}
  {\bfseries A904-905} (2013) 409c--412c},
\href{http://arxiv.org/abs/1210.5144}{{\ttfamily arXiv:1210.5144 [hep-ph]}}.

\bibitem{Alver:2010gr}
B.~Alver and G.~Roland, ``{Collision geometry fluctuations and triangular flow
  in heavy-ion collisions},''
  \href{http://dx.doi.org/10.1103/PhysRevC.82.039903,
  10.1103/PhysRevC.81.054905}{{\em Phys.Rev.} {\bfseries C81} (2010) 054905},
\href{http://arxiv.org/abs/1003.0194}{{\ttfamily arXiv:1003.0194 [nucl-th]}}.

\bibitem{McLerran:1993ni}
L.~D. McLerran and R.~Venugopalan, ``{Computing quark and gluon distribution
  functions for very large nuclei},'' {\em Phys. Rev.} {\bfseries D49} (1994)
  2233--2241.

\bibitem{McLerran:1993ka}
L.~D. McLerran and R.~Venugopalan, ``{Gluon distribution functions for very
  large nuclei at small transverse momentum},'' {\em Phys. Rev.} {\bfseries
  D49} (1994) 3352--3355.

\bibitem{GolecBiernat:1998js}
K.~J. Golec-Biernat and M.~Wusthoff, ``{Saturation effects in deep inelastic
  scattering at low $Q^2$ and its implications on diffraction},'' {\em Phys.
  Rev.} {\bfseries D59} (1999) 014017.

\bibitem{GolecBiernat:1999qd}
K.~J. Golec-Biernat and M.~Wusthoff, ``{Saturation in diffractive deep
  inelastic scattering},'' {\em Phys. Rev.} {\bfseries D60} (1999) 114023.

\bibitem{Broniowski:2007nz}
W.~Broniowski, M.~Rybczynski, and P.~Bozek, ``{GLISSANDO: Glauber initial-state
  simulation and more..},''
  \href{http://dx.doi.org/10.1016/j.cpc.2008.07.016}{{\em Comput.Phys.Commun.}
  {\bfseries 180} (2009) 69--83},
\href{http://arxiv.org/abs/0710.5731}{{\ttfamily arXiv:0710.5731 [nucl-th]}}.

\bibitem{Rybczynski:2013yba}
M.~Rybczynski, G.~Stefanek, W.~Broniowski, and P.~Bozek, ``{GLISSANDO 2:
  GLauber Initial-State Simulation AND mOre..., ver. 2},''
  \href{http://dx.doi.org/10.1016/j.cpc.2014.02.016}{{\em Comput.Phys.Commun.}
  {\bfseries 185} (2014) 1759--1772},
\href{http://arxiv.org/abs/1310.5475}{{\ttfamily arXiv:1310.5475 [nucl-th]}}.

\bibitem{Bialas:1976ed}
A.~Bialas, M.~Bleszynski, and W.~Czyz, ``{Multiplicity Distributions in
  Nucleus-Nucleus Collisions at High-Energies},''
\href{http://dx.doi.org/10.1016/0550-3213(76)90329-1}{{\em Nucl.Phys.}
  {\bfseries B111} (1976) 461}.

\bibitem{Landau:1953gs}
L.~D. Landau, ``On the multiparticle production in high-energy collisions,''
{\em Izv. Akad. Nauk SSSR Ser. Fiz.} {\bfseries 17} (1953) 51--64.

\bibitem{Khalatnikov:1954}
I.~M. Khalatnikov, ``Some questions of the relativistic hydrodynamics,'' {\em
  Zh.Eksp. Teor. Fiz. .} {\bfseries 26} (1954) 529.

\bibitem{Belenkij:1956cd}
S.~Z. Belenkij and L.~D. Landau, ``Hydrodynamic theory of multiple production
  of particles,'' {\em Nuovo Cim. Suppl.} {\bfseries 3S10} (1956) 309.

\bibitem{Koppe:1948}
H.~Koppe {\em Z. f. Naturforsch.} {\bfseries 3A} (1948) 251.

\bibitem{Fermi:1950jd}
E.~Fermi, ``High-energy nuclear events,''
{\em Prog. Theor. Phys.} {\bfseries 5} (1950) 570--583.

\bibitem{Fermi:1951zz}
E.~Fermi, ``{Angular Distribution of the Pions Produced in High Energy Nuclear
  Collisions},''
{\em Phys. Rev.} {\bfseries 81} (1951) 683--687.

\bibitem{Pomeranchuk:1951ey}
I.~Y. Pomeranchuk, ``On the theory of multiple particle production in a single
  collision,''
{\em Dokl. Akad. Nauk Ser. Fiz.} {\bfseries 78} (1951) 889--891.

\bibitem{Bjorken:1982qr}
J.~Bjorken, ``{Highly Relativistic Nucleus-Nucleus Collisions: The Central
  Rapidity Region},''
\href{http://dx.doi.org/10.1103/PhysRevD.27.140}{{\em Phys.Rev.} {\bfseries
  D27} (1983) 140--151}.

\bibitem{Baym:1983sr}
G.~Baym, B.~L. Friman, J.~P. Blaizot, M.~Soyeur, and W.~Czyz, ``Hydrodynamics
  of ultrarelativistic heavy ion collisions,''
{\em Nucl. Phys.} {\bfseries A407} (1983) 541--570.

\bibitem{Haque:2013sja}
N.~Haque, J.~O. Andersen, M.~G. Mustafa, M.~Strickland, and N.~Su,
  ``{Three-loop HTLpt Pressure and Susceptibilities at Finite Temperature and
  Density},'' \href{http://dx.doi.org/10.1103/PhysRevD.89.061701}{{\em
  Phys.Rev.} {\bfseries D89} (2014) 061701},
\href{http://arxiv.org/abs/1309.3968}{{\ttfamily arXiv:1309.3968 [hep-ph]}}.

\bibitem{Haque:2014rua}
N.~Haque, A.~Bandyopadhyay, J.~O. Andersen, M.~G. Mustafa, M.~Strickland, {\em
  et~al.}, ``{Three-loop HTLpt thermodynamics at finite temperature and
  chemical potential},'' \href{http://dx.doi.org/10.1007/JHEP05(2014)027}{{\em
  JHEP} {\bfseries 1405} (2014) 027},
\href{http://arxiv.org/abs/1402.6907}{{\ttfamily arXiv:1402.6907 [hep-ph]}}.

\bibitem{Borsanyi:2010cj}
S.~Borsanyi, G.~Endrodi, Z.~Fodor, A.~Jakovac, S.~D. Katz, {\em et~al.}, ``{The
  QCD equation of state with dynamical quarks},''
  \href{http://dx.doi.org/10.1007/JHEP11(2010)077}{{\em JHEP} {\bfseries 1011}
  (2010) 077},
\href{http://arxiv.org/abs/1007.2580}{{\ttfamily arXiv:1007.2580 [hep-lat]}}.

\bibitem{Bhattacharya:2014ara}
T.~Bhattacharya, M.~I. Buchoff, N.~H. Christ, H.~T. Ding, R.~Gupta, {\em
  et~al.}, ``{The QCD phase transition with physical-mass, chiral quarks},''
  \href{http://dx.doi.org/10.1103/PhysRevLett.113.082001}{{\em Phys.Rev.Lett.}
  {\bfseries 113} (2014) 082001},
\href{http://arxiv.org/abs/1402.5175}{{\ttfamily arXiv:1402.5175 [hep-lat]}}.

\bibitem{Chojnacki:2007jc}
M.~Chojnacki and W.~Florkowski, ``{Temperature dependence of sound velocity and
  hydrodynamics of ultra-relativistic heavy-ion collisions},'' {\em Acta
  Phys.Polon.} {\bfseries B38} (2007) 3249--3262,
\href{http://arxiv.org/abs/nucl-th/0702030}{{\ttfamily arXiv:nucl-th/0702030
  [NUCL-TH]}}.

\bibitem{Broniowski:2008vp}
W.~Broniowski, M.~Chojnacki, W.~Florkowski, and A.~Kisiel, ``{Uniform
  Description of Soft Observables in Heavy-Ion Collisions at s(NN)**(1/2) = 200
  GeV**2},'' \href{http://dx.doi.org/10.1103/PhysRevLett.101.022301}{{\em
  Phys.Rev.Lett.} {\bfseries 101} (2008) 022301},
\href{http://arxiv.org/abs/0801.4361}{{\ttfamily arXiv:0801.4361 [nucl-th]}}.

\bibitem{Florkowski:2010mc}
W.~Florkowski, ``{The realistic QCD equation of state in relativistic heavy-ion
  collisions and the early Universe},''
  \href{http://dx.doi.org/10.1016/j.nuclphysa.2011.01.024}{{\em Nucl.Phys.}
  {\bfseries A853} (2011) 173--188},
\href{http://arxiv.org/abs/1008.5225}{{\ttfamily arXiv:1008.5225 [nucl-th]}}.

\bibitem{Wang:2000bf}
X.-N. Wang and M.~Gyulassy, ``{Energy and centrality dependence of rapidity
  densities at RHIC},''
  \href{http://dx.doi.org/10.1103/PhysRevLett.86.3496}{{\em Phys.Rev.Lett.}
  {\bfseries 86} (2001) 3496--3499},
\href{http://arxiv.org/abs/nucl-th/0008014}{{\ttfamily arXiv:nucl-th/0008014
  [nucl-th]}}.

\bibitem{Kharzeev:2000ph}
D.~Kharzeev and M.~Nardi, ``{Hadron production in nuclear collisions at RHIC
  and high density QCD},''
  \href{http://dx.doi.org/10.1016/S0370-2693(01)00457-9}{{\em Phys.Lett.}
  {\bfseries B507} (2001) 121--128},
\href{http://arxiv.org/abs/nucl-th/0012025}{{\ttfamily arXiv:nucl-th/0012025
  [nucl-th]}}.

\bibitem{Israel:1979wp}
W.~Israel and J.~Stewart, ``{Transient relativistic thermodynamics and kinetic
  theory},''
\href{http://dx.doi.org/10.1016/0003-4916(79)90130-1}{{\em Annals Phys.}
  {\bfseries 118} (1979) 341--372}.

\bibitem{Noronha-Hostler:2013gga}
J.~Noronha-Hostler, G.~S. Denicol, J.~Noronha, R.~P.~G. Andrade, and F.~Grassi,
  ``{Bulk Viscosity Effects in Event-by-Event Relativistic Hydrodynamics},''
  \href{http://dx.doi.org/10.1103/PhysRevC.88.044916}{{\em Phys.Rev.}
  {\bfseries C88} (2013) 044916},
\href{http://arxiv.org/abs/1305.1981}{{\ttfamily arXiv:1305.1981 [nucl-th]}}.

\bibitem{Strickland:2014pga}
M.~Strickland, ``{Anisotropic Hydrodynamics: Three lectures},''
\href{http://arxiv.org/abs/1410.5786}{{\ttfamily arXiv:1410.5786 [nucl-th]}}.

\bibitem{Florkowski:2010cf}
W.~Florkowski and R.~Ryblewski, ``{Highly-anisotropic and strongly-dissipative
  hydrodynamics for early stages of relativistic heavy-ion collisions},''
  \href{http://dx.doi.org/10.1103/PhysRevC.83.034907}{{\em Phys.Rev.}
  {\bfseries C83} (2011) 034907},
\href{http://arxiv.org/abs/1007.0130}{{\ttfamily arXiv:1007.0130 [nucl-th]}}.

\bibitem{Martinez:2010sc}
M.~Martinez and M.~Strickland, ``{Dissipative Dynamics of Highly Anisotropic
  Systems},'' \href{http://dx.doi.org/10.1016/j.nuclphysa.2010.08.011}{{\em
  Nucl.Phys.} {\bfseries A848} (2010) 183--197},
\href{http://arxiv.org/abs/1007.0889}{{\ttfamily arXiv:1007.0889 [nucl-th]}}.

\bibitem{Hung:1997du}
C.~Hung and E.~V. Shuryak, ``{Equation of state, radial flow and freezeout in
  high-energy heavy ion collisions},''
  \href{http://dx.doi.org/10.1103/PhysRevC.57.1891}{{\em Phys.Rev.} {\bfseries
  C57} (1998) 1891--1906},
\href{http://arxiv.org/abs/hep-ph/9709264}{{\ttfamily arXiv:hep-ph/9709264
  [hep-ph]}}.

\bibitem{Koch:1985hk}
P.~Koch and J.~Rafelski, ``{Why the Hadronic Gas Description of Hadronic
  Reactions Works: The Example of Strange Hadrons},''
{\em South Afr.J.Phys.} {\bfseries 9} (1986) 8.

\bibitem{Cleymans:1992zc}
J.~Cleymans and H.~Satz, ``{Thermal hadron production in high-energy heavy ion
  collisions},'' \href{http://dx.doi.org/10.1007/BF01555746}{{\em Z.Phys.}
  {\bfseries C57} (1993) 135--148},
\href{http://arxiv.org/abs/hep-ph/9207204}{{\ttfamily arXiv:hep-ph/9207204
  [hep-ph]}}.

\bibitem{Gazdzicki:1998vd}
M.~Gazdzicki and M.~I. Gorenstein, ``{On the early stage of nucleus-nucleus
  collisions},'' {\em Acta Phys.Polon.} {\bfseries B30} (1999) 2705,
\href{http://arxiv.org/abs/hep-ph/9803462}{{\ttfamily arXiv:hep-ph/9803462
  [hep-ph]}}.

\bibitem{Becattini:2000jw}
F.~Becattini, J.~Cleymans, A.~Keranen, E.~Suhonen, and K.~Redlich, ``{Features
  of particle multiplicities and strangeness production in central heavy ion
  collisions between 1.7A-GeV/c and 158A-GeV/c},''
  \href{http://dx.doi.org/10.1103/PhysRevC.64.024901}{{\em Phys.Rev.}
  {\bfseries C64} (2001) 024901},
\href{http://arxiv.org/abs/hep-ph/0002267}{{\ttfamily arXiv:hep-ph/0002267
  [hep-ph]}}.

\bibitem{BraunMunzinger:2001ip}
P.~Braun-Munzinger, D.~Magestro, K.~Redlich, and J.~Stachel, ``{Hadron
  production in Au - Au collisions at RHIC},''
  \href{http://dx.doi.org/10.1016/S0370-2693(01)01069-3}{{\em Phys.Lett.}
  {\bfseries B518} (2001) 41--46},
\href{http://arxiv.org/abs/hep-ph/0105229}{{\ttfamily arXiv:hep-ph/0105229
  [hep-ph]}}.

\bibitem{Florkowski:2001fp}
W.~Florkowski, W.~Broniowski, and M.~Michalec, ``{Thermal analysis of particle
  ratios and p(t) spectra at RHIC},'' {\em Acta Phys.Polon.} {\bfseries B33}
  (2002) 761--769,
\href{http://arxiv.org/abs/nucl-th/0106009}{{\ttfamily arXiv:nucl-th/0106009
  [nucl-th]}}.

\bibitem{Torrieri:2004zz}
G.~Torrieri, S.~Steinke, W.~Broniowski, W.~Florkowski, J.~Letessier, {\em
  et~al.}, ``{SHARE: Statistical hadronization with resonances},''
  \href{http://dx.doi.org/10.1016/j.cpc.2005.01.004}{{\em Comput.Phys.Commun.}
  {\bfseries 167} (2005) 229--251},
\href{http://arxiv.org/abs/nucl-th/0404083}{{\ttfamily arXiv:nucl-th/0404083
  [nucl-th]}}.

\bibitem{Petran:2013lja}
M.~Petran, J.~Letessier, V.~Petracek, and J.~Rafelski, ``{Hadron production and
  quark-gluon plasma hadronization in Pb-Pb collisions at $\sqrt{s_{NN}}=2.76$
  TeV},'' \href{http://dx.doi.org/10.1103/PhysRevC.88.034907}{{\em Phys.Rev.}
  {\bfseries C88} no.~3, (2013) 034907},
\href{http://arxiv.org/abs/1303.2098}{{\ttfamily arXiv:1303.2098 [hep-ph]}}.

\bibitem{Stachel:2013zma}
J.~Stachel, A.~Andronic, P.~Braun-Munzinger, and K.~Redlich, ``{Confronting LHC
  data with the statistical hadronization model},''
  \href{http://dx.doi.org/10.1088/1742-6596/509/1/012019}{{\em
  J.Phys.Conf.Ser.} {\bfseries 509} (2014) 012019},
\href{http://arxiv.org/abs/1311.4662}{{\ttfamily arXiv:1311.4662 [nucl-th]}}.

\bibitem{Floris:2014pta}
M.~Floris, ``{Hadron yields and the phase diagram of strongly interacting
  matter},''
\href{http://arxiv.org/abs/1408.6403}{{\ttfamily arXiv:1408.6403 [nucl-ex]}}.

\bibitem{Broniowski:2001we}
W.~Broniowski and W.~Florkowski, ``{Explanation of the RHIC p(T) spectra in a
  thermal model with expansion},''
  \href{http://dx.doi.org/10.1103/PhysRevLett.87.272302}{{\em Phys.Rev.Lett.}
  {\bfseries 87} (2001) 272302},
\href{http://arxiv.org/abs/nucl-th/0106050}{{\ttfamily arXiv:nucl-th/0106050
  [nucl-th]}}.

\bibitem{Kisiel:2005hn}
A.~Kisiel, T.~Taluc, W.~Broniowski, and W.~Florkowski, ``{THERMINATOR: THERMal
  heavy-IoN generATOR},''
  \href{http://dx.doi.org/10.1016/j.cpc.2005.11.010}{{\em Comput.Phys.Commun.}
  {\bfseries 174} (2006) 669--687},
\href{http://arxiv.org/abs/nucl-th/0504047}{{\ttfamily arXiv:nucl-th/0504047
  [nucl-th]}}.

\bibitem{Chojnacki:2011hb}
M.~Chojnacki, A.~Kisiel, W.~Florkowski, and W.~Broniowski, ``{THERMINATOR 2:
  THERMal heavy IoN generATOR 2},''
  \href{http://dx.doi.org/10.1016/j.cpc.2011.11.018}{{\em Comput.Phys.Commun.}
  {\bfseries 183} (2012) 746--773},
\href{http://arxiv.org/abs/1102.0273}{{\ttfamily arXiv:1102.0273 [nucl-th]}}.

\bibitem{Abelev:2012wca}
{\bfseries ALICE Collaboration} Collaboration, B.~Abelev {\em et~al.}, ``{Pion,
  Kaon, and Proton Production in Central Pb--Pb Collisions at $\sqrt{s_{NN}} =
  2.76$ TeV},'' \href{http://dx.doi.org/10.1103/PhysRevLett.109.252301}{{\em
  Phys.Rev.Lett.} {\bfseries 109} (2012) 252301},
\href{http://arxiv.org/abs/1208.1974}{{\ttfamily arXiv:1208.1974 [hep-ex]}}.

\bibitem{Abelev:2013vea}
{\bfseries ALICE Collaboration} Collaboration, B.~Abelev {\em et~al.},
  ``{Centrality dependence of $\pi$, K, p production in Pb-Pb collisions at
  $\sqrt{s_{NN}}$ = 2.76 TeV},''
  \href{http://dx.doi.org/10.1103/PhysRevC.88.044910}{{\em Phys.Rev.}
  {\bfseries C88} no.~4, (2013) 044910},
\href{http://arxiv.org/abs/1303.0737}{{\ttfamily arXiv:1303.0737 [hep-ex]}}.

\bibitem{Begun:2013nga}
V.~Begun, W.~Florkowski, and M.~Rybczynski, ``{Explanation of hadron
  transverse-momentum spectra in heavy-ion collisions at sqrt(sNN) = 2.76 TeV
  within chemical non-equilibrium statistical hadronization model},''
  \href{http://dx.doi.org/10.1103/PhysRevC.90.014906}{{\em Phys.Rev.}
  {\bfseries C90} (2014) 014906},
\href{http://arxiv.org/abs/1312.1487}{{\ttfamily arXiv:1312.1487 [nucl-th]}}.

\bibitem{Begun:2014rsa}
V.~Begun, W.~Florkowski, and M.~Rybczynski, ``{Transverse-momentum spectra of
  strange particles produced in Pb+Pb collisions at $\sqrt{s_{\rm NN}}=2.76$
  TeV in the chemical non-equilibrium model},''
\href{http://arxiv.org/abs/1405.7252}{{\ttfamily arXiv:1405.7252 [hep-ph]}}.

\bibitem{Petersen:2008dd}
H.~Petersen, J.~Steinheimer, G.~Burau, M.~Bleicher, and H.~Stocker, ``{A Fully
  Integrated Transport Approach to Heavy Ion Reactions with an Intermediate
  Hydrodynamic Stage},''
  \href{http://dx.doi.org/10.1103/PhysRevC.78.044901}{{\em Phys.Rev.}
  {\bfseries C78} (2008) 044901},
\href{http://arxiv.org/abs/0806.1695}{{\ttfamily arXiv:0806.1695 [nucl-th]}}.

\end{thebibliography}\endgroup


\end{document}